\documentclass[twocolumn,showpacs,preprintnumbers,amsmath,amssymb,floatfix]{revtex4}

\usepackage{graphicx}
\usepackage{dcolumn}
\usepackage{bm}

\begin{document}

\title{
Exact Thermodynamics of Pairing and Charge-spin Separation\\
Crossovers in Small Hubbard Nanoclusters}

\author{Armen~N.~Kocharian} 

\affiliation{ Department of Physics and Astronomy, California
State University, Northridge, \\ CA 91330-8268 }

\author{Gayanath~W.~Fernando}
\affiliation{Department of Physics, University of Connecticut,\\
 Storrs, CT 06269 and IFS, Hantana Rd., Kandy, Sri Lanka}

\author{Tun Wang}

\address{Department of Physics, University of Connecticut,\\
 Storrs, CT 06269}
\author{Kalum~Palandage}
\affiliation{Department of Physics, University of Connecticut,\\
 Storrs, CT 06269}

\author{James~W.~Davenport}
\affiliation{Computational Science Center, Brookhaven National
Laboratory,\\
Upton, NY 11973}

\begin{abstract}
The exact numerical diagonalization and thermodynamics in an
ensemble of small Hubbard clusters in the ground state and finite
temperatures reveal intriguing insights into the nascent charge
and spin pairings, Bose condensation and ferromagnetism in
nanoclusters. The phase diagram off half filling strongly suggests
the existence of subsequent transitions from electron pairing into
unsaturated and saturated ferromagnetic Mott-Hubbard like
insulators, driven by electron repulsion. Rigorous criteria for 
the existence of quantum critical points in the ground state and
corresponding crossovers at finite temperatures are formulated.
The phase diagram for ${2\times 4}$-site clusters illustrates how
these features are scaled with cluster size. The phase separation
and electron pairing, monitored by a magnetic field and electron
doping, surprisingly resemble phase diagrams in 
the family of doped high T$_c$ cuprates.
\end{abstract}

\maketitle
\section{Introduction}
Despite tremendous experimental and theoretical efforts, there is
still no microscopic theory that can yield  comprehensive support
for the bare Coulomb interaction originated pairing correlations,
phase separation and pseudogap phenomena in clusters, small
nanoparticles, transition metal oxides and high T$_c$ cuprates
~\cite{deHeer0,deHeer1,Tinkham,Cox,RVB,Nature,Timusk,Kivelson_Review,Marshall,hashini}.
The recent discovery of the ferromagnetic insulators at
room-temperature has further stimulated a great interest 
related to the role of on-site Coulomb interaction in the origin
of ferromagnetism~\cite{Yuji}. Electrons in a finite Hubbard
lattice, subjected to strong on-site electron repulsion near half
filling, can lead to  spontaneous ferromagnetism~\cite{Nagaoka}
and the finite temperature phase diagram is expected to be
applicable to disulfides~\cite{Sokoloff}. Moderate Coulomb
interaction can also lead to phase separation and formation of
mesoscale structures (such as "stripes") under doping of Mott
insulating "parent" materials with highly correlated electrons,
including high T$_c$ cuprates~\cite{Nagaev,Tallon,Andrea,Zachar}.
Although the experimental determination of various inhomogeneous
phases in the cuprates is still somewhat
controversial~\cite{Tallon,Andrea}, the underdoped high T$_c$
superconductors (HTSCs) have many common features and are often
characterized by crossover temperatures below which excitation
(pseudo)gaps in the normal-state are seen to
develop~\cite{Zachar}. The detailed manner in which T$_c$ and
crossover temperature changes under variation of electron
concentration, magnetic field or pressure (Coulomb interaction) is
also of fundamental interest for the formulation of the
microscopic models responsible for nascent
superconductivity~\cite{Schilling}. In the optimally doped
cuprates, the correlation length of dynamical spin fluctuations is
very small~\cite{Zha} and 
 hence short-range fluctuations are
dominant over long-range ones. Therefore, a microscopic theory,
with short-range dynamical correlations, can give useful insight
into nascent superconductivity in clusters and the rich physics
observed in the high-T$_c$ cuprates. In our opinion, thermodynamic
properties of small Hubbard clusters under variation of
composition, size, structure, temperature and magnetic field have
not been fully explored, although there have been numerous exact
calculations~\cite{Shiba,schumann}. From this perspective the
exact diagonalization of small Hubbard {\it nanoclusters} can give
insights related to
 the origin of superconductivity and
ferromagnetism in an ensemble of clusters, nanoparticles, and
eventually, nanomaterials~\cite{JMMM,PRB}.

The following questions are central to our study: (i) Using exact
cluster studies, is it possible to obtain a microscopic
understanding of charge and spin separation and electron charge
pairing and identify various possible incipient phases at finite
temperatures? (ii) Is it possible for ferromagnetism to occur in
Mott-Hubbard like insulators away from half filling? (iii) Do
these {\it nanocluster} phase diagrams retain important features,
such as quantum critical points and crossovers that are known for
mesoscopic structures and large thermodynamic systems? (iv) When
treated exactly, what essential features can the simple Hubbard
clusters capture that are in common with the transition metal
oxides, cuprates, etc.?

In addition, 4-site (square) cluster is the basic building block
of the CuO$_2$ planes in the HTSCs and it can be used as a block
reference to build up larger superblocks in 2D of desirable
$L\times L$ sizes~\cite{CPT,tk,rsrg,CRG,Jarrell}. Short range
electronic correlations provide unique insight into the Nagaoka
ferromagnetism in small 2D and 3D ferromagnetic
particles~\cite{Nagaoka,Sokoloff}, exact thermodynamics of
many-body physics, which are difficult to obtain from approximate
methods~\cite{JMMM}. Here we present also strong evidence
underlining the occurrence of saturated and unsaturated
ferromagnetism~\cite{Sebold}, particle-particle, particle-hole
{\it pairings} and corresponding temperature driven Bose
condensation (BC) {\it crossovers} for spin and charge degrees in
mesoscale structures.

The paper is organized as follows. In the following section we
present the model and formulate exact thermodynamics in grand
canonical ensemble approach. In the third section, we introduce
the charge and spin pseudogaps and define corresponding new order
parameters. In section four, we present the results of numerical
calculations for electron density and magnetization versus
chemical potential and illustrate how to calculate
 various phase boundaries for particle-particle/hole
pairing, phase separation instabilities, quantum critical points
in the ground state and crossovers at finite temperatures. The
results for the ${2\times4}$ cluster are used to illustrate how
these features are scaled with the cluster size. The concluding
summary is presented in the closing section.

\section{Model and formalism}

The quantum and thermal fluctuations of electrons in finite
clusters can be described by the Hubbard Hamiltonian, placed in a
magnetic field $h$ as,
\begin{eqnarray}
H=-t\sum\limits_{<ij, \sigma>
\sigma}c^{+}_{i\sigma}~c_{j\sigma}-\mu\sum\limits_{i, \sigma
}n_{i\sigma} +\sum\limits_{i} U n_{i\uparrow}
n_{i\downarrow}-\nonumber\\ h \gamma\sum\limits_{i \sigma}
(n_{i\uparrow}-n_{i\downarrow}),\label{2-site1}
\end{eqnarray}
with the hopping amplitude $t$ (energies are measured in the units
of $t$) and on-site Coulomb interaction
 $U\ge 0$. Here $\gamma={1\over 2}$ is the magnetic moment of an electron and $\mu$ is
 the chemical potential for the ensemble of clusters.
 This work utilizes {\it statistical} canonical
 and grand canonical ensembles using analytical eigenvalues for 4-site
 clusters with periodic boundary conditions~\cite{PRB}.
 In addition, here we present the results of exact numerical diagonalization
 of $2\times 4$ square 2D clusters,
 using numerical eigenvalues
 in the above ensembles to study thermal and quantum {\sl
 fluctuations}.

In nanoparticles electrons and holes are limited to small regions
and an effect known as quantum confinement yields discrete
spectrum and energy-level spacings between filled and empty states
that can modify the thermodynamics. Because the small clusters are
far from thermodynamic limit, one can naively think that the
standard tools for the description of phase transitions are not
applicable and other concepts are needed. However, it is important
to realize that phase transitions and corresponding temperature
driven crossovers in the grand canonical ensemble can very well be
defined and classified for finite systems without the use of the
thermodynamic limit. It is further shown also how spatially
inhomogeneous configurations like phase separations can be
obtained using the canonical ensemble. We illustrate how phase
transitions on the verge of an instability and phase separation
(segregation) can be defined and classified unambiguously in
finite Hubbard clusters~\cite{PRB}. It is thus possible to define
phase transitions and crossovers even in small systems with local
Coulomb forces, which are not "thermodynamically stable". There is
also strong support with regard to the effectiveness of the grand
canonical approach for studies of magnetism in small clusters due
to recent experimental findings that average magnetization of a
ferromagnetic cluster is a property of the ensemble of isolated
clusters but not of the individual cluster~\cite{deHeer0,deHeer1}.

One can eliminate next nearest neighbor couplings by replacing the
planar square lattice with independent 4-site clusters and treat
them as an ensemble immersed in a particle heat reservoir, where
electrons can transfer from cluster to cluster due to the thermal
fluctuations~\cite{PRB}. Earlier we formulated a new scheme in
which thermodynamical/statistical notions, concepts and
theoretical methodologies are tailored for the calculations of
exact thermodynamics in a grand canonical ensemble of 4-site
clusters. Degrees of freedom for charge and spin, electron and
spin pairings, temperature crossovers, quantum critical points,
etc. were extracted directly from the thermodynamics of these
clusters~\cite{JMMM,PRB}. The grand partition function $Z$ (where
the number of particles $N$ and the projection of spin $s^z$ are
allowed to fluctuate) and its derivatives are calculated exactly
without taking the thermodynamic limit~\cite{JMMM}. The charge and
spin fluctuation responses to electron or hole doping level (i.e.
chemical potential $\mu$) and an applied magnetic field ($h$)
resulting in weak {\it saddle point} singularities $({\it
crossovers})$, which display clearly identifiable, prominent peaks
in corresponding charge and spin density of states~\cite{PRB}.

\section{Charge and spin pseudogaps}

We apply the grand canonical ensemble of decoupled clusters in
contact with a bath reservoir allowing for the particle number to
fluctuate. It is straightforward to calculate the above
thermodynamic properties and some of these results for the 2- and
4-site clusters were reported earlier~\cite{JMMM}. Using these
eigenvalues, we have evaluated the {\sl exact} grand partition
function and thermal averages such as magnetization and
susceptibilities numerically as a function of the set of
parameters $\{T,h,\mu,U\}$. Using peaks in spin and charge
susceptibilities, phase diagrams in a $T$ vs $\mu$ plane for
arbitrary $U$ and $h$ can be constructed. This approach also
allows us to obtain quantum critical points (QCPs) and rigorous
criteria for various sharp transitions, such as the Mott-Hubbard
(MH), antiferromagnetic (AF) or ferromagnetic (F) transitions in
the ground state~\cite{JMMM} and charge (particle-particle) or
spin condensation at finite temperatures~\cite{PRB} using peaks in
charge or spin susceptibilities (see below).

\begin{figure} 
\begin{center}
\includegraphics*[width=20pc,height=20pc]{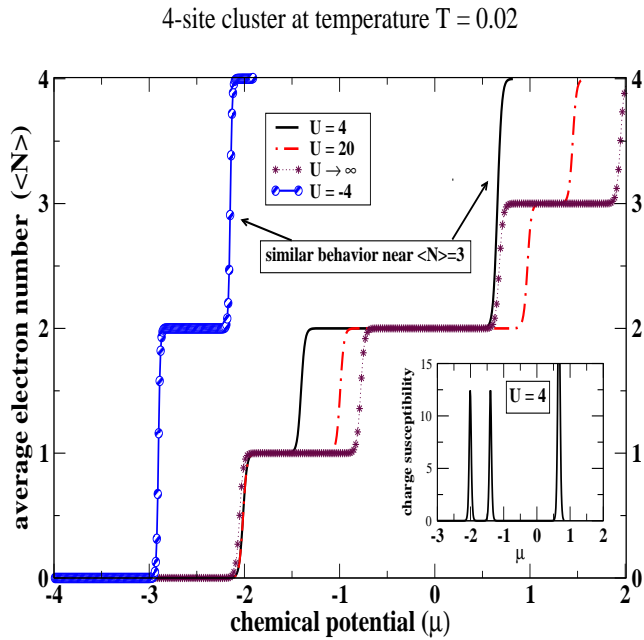}
\end{center}
\caption {Variation of average electron concentration versus $\mu$
in ensemble of 4-site clusters for various $U$ values. The result
for $U=-4$ is given for comparison with $U=4$. The inset shows the
variation of charge susceptibility $\chi_c$ versus $\mu$ for
$U=4$.} \label{fig:num}
\end{figure}

The difference in energies for a given temperature between
configurations with various numbers of electrons is obtained by
adding or subtracting one electron (charge) and defined as,
\begin{eqnarray}
{\Delta^{c}}(T) = [E(M-1,M^\prime;U,T)-E(M+1,M^\prime;U,T)]-\nonumber \\
2[E(M,M^\prime;U,T)-E(M+1,M^\prime;U,T)], \label{charge_gap}
\end{eqnarray}
where $E(M,M^\prime;U,T)$ is the lowest canonical (ensemble)
energy with a given number of electrons $N=M+M^{\prime}$
determined by the number of up ($M$) and down ($M^\prime$) spins.
The quantity ${\Delta^{c}}(T)$ is related to the discretized
second derivative of the energy with respect to the number of
particles, i.e. charge susceptibility $\chi_c$. We define the
chemical potential energies at finite temperature as $\mu_{\pm}$,
\begin{eqnarray}
&\mu_{+}=E(M+1,M^\prime;U:T)-E(M,M^\prime;U,T)\label{mupm1}  \\
&\mu_{-}=E(M,M^\prime;U:T)-E(M-1,M^\prime;U:T). \label{mupm2}
\end{eqnarray}
Eqs. (\ref{mupm1}) and (\ref{mupm2}) at $T=0$ are identical to
those introduced in~\cite{Lieb1}, near half filling. One usually
refers to MH transitions, plateaus or gaps in $\left\langle
N\right\rangle$ vs $\mu$, as those occurring only at half filling
$\left\langle N\right\rangle=4$ in the 4-site cluster with lower
and upper Hubbard subbands separated by the energy gap. At half
filling, one can recall MH and AF critical temperatures, at which
corresponding MH and AF gaps disappear. Similar steps (gaps) at
other $\left\langle N\right\rangle$ in charge and spin degrees
will be referred to as MH-like and AF-like plateaus respectively
in order to distinguish these.
 This terminology would also apply
to all the labelings in Fig.~\ref{fig:num} and the rest of the
paper. Using the definitions (\ref{mupm1}) and (\ref{mupm2}), the
corresponding charge energy gap, ${\Delta^{c}}(T)$, at finite
temperature can be written as a difference, ${\Delta^{c}}(T) =
{\mu_{+}} - {\mu_{-}}$. Notice that $\mu_{+}(T)$ and $\mu_{-}(T)$
identify  peak positions in
a  $T-\mu$ space for charge susceptibility $\chi_c$  at finite
temperature. The condition ${\Delta^{c}}(T)>0$ provides
electron-hole (excitonic) excitations, with a positive pseudogap,
$\Delta^{e-h}(T)\ge 0$~\cite{JMMM}. Accordingly, the condition
${\Delta^{c}}(T)<0$ with $\mu_{-}(T)>\mu_{+}(T)$ gives
electron-electron pairing with positive energy,
${\Delta^{P}}(T)>0$~\cite{PRB}. Using the above definitions for
$\mu_{\pm}$ we can combine 
them and write:
\begin{eqnarray}
\Delta^{c}(T) = \left\{ \begin{array}{ll}
\Delta^{e-h}(T), & \mbox{for $\mu_{+} > \mu_{-}$} \\
\Delta^{P}(T), & \mbox{for $\mu_{-} > \mu_{+}$}, \\
\end{array}
\right. \label{electron-hole_gap}
\end{eqnarray}
where $\Delta^{e-h}(T)\ge 0$ serves as a natural order parameter
and will be called a MH-like {\sl pseudogap} at nonzero
temperature, since $\chi$ has a small, but nonzero weight inside
the gap at infinitesimal temperature. At $T=0$, this gap will be
labeled  a true gap since $\chi_c$ is exactly zero inside. For a
given $U$, we define the crossover critical temperature $T_c(U)$
at which the electron-hole pairing pseudogap vanishes and a Fermi
liquid state, with $\mu_{-} = \mu_{+}$, 
 becomes stable.

At zero temperature, the expression for electron binding energy
$\Delta^{P}(T)$ is identical to {\it true gap} introduced in
Ref.~\cite{white} and at nonzero temperature it will be called a
pairing pseudogap. We can easily trace the peaks of $\chi_c(\mu)$
at {\it finite} temperature, and find $T_c(\mu)$ at which a
possible maximum occurs for a given $\mu$. At finite temperature,
the charge susceptibility is a differentiable function of $\mu$
and the peaks, {\sl which may exist in a limited range of
temperature}, are identified easily from the conditions,
$\chi^{'}_c(\mu_{\pm})=0$ with $\chi^{''}_c(\mu_{\pm}) <0$. The
charge pseudogap at half filling $\mu_0= \mu_+ =\mu_-={U\over 2}$
vanishes at $T_{MH}$ which can be identified clearly with
$\chi^{'}_c(\mu_0) =0$ and $\chi^{''}_c(\mu_0) =0$; i.e. as the
temperature corresponding to a point of inflexion in
$\chi_c(\mu)$. This procedure describes rigorous conditions for
identifying the MH transition temperature and a similar one can be
carried out for the spin gap, $\Delta^s(T)$, by following the
evolution of spin susceptibilities $\chi_s(\mu)$ as a function of
$\mu$. Possible peaks in the zero magnetic field spin
susceptibility $\chi_s(\mu)$, when monitored as a function of
$\mu$, can be used to define an associated temperature,
$T_s(\mu)$, as the temperature at which such a peak exists and a
spin pseudogap as the separation between two such peaks.

Similar to the charge plateaus seen in $\left\langle
N\right\rangle$ {\it versus} $\mu$, we can trace the variation of
magnetization $\left\langle s^z\right\rangle$ {\it versus} an
applied magnetic field $h$ and identify the spin plateau features
which can be associated with staggered magnetization or short
range AF. We calculate the critical magnetic field $h_{c\pm}$ for
the onset of magnetization which depends on $U$, $T$ and $N$, by
flipping a down spin, or an up spin~\cite{PRB}. The difference in
energies for given temperature between configurations with various
spins defines the spin pseudogap
\begin{eqnarray}
{\Delta^{s}}(T) = {{h_{+}} - {h_{-}}\over 2}, \label{spin_gap}
\end{eqnarray}
where $h_{+}$ and $h_{-}$ identify the peak positions in the $h$
space for the spin susceptibility $\chi_s$. The condition
${\Delta^{s}}(T)>0$
yields spin (AF-like) paring, $\Delta^{AF}(T)>0$, with a positive
(spin) pseudogap~\cite{PRB}. Accordingly, the condition
${\Delta^{s}}(T)<0$ with $h_{-}(T)>h_{+}(T)$
yields a ferromagnetic pairing pseudogap, $\Delta^{F}(T)>0$, for
spin (triplet or quadruplet) coupling. Using the above definitions
for $h_{\pm}$ we can combine both as,
\begin{eqnarray}
\Delta^{s}(T) = \left\{ \begin{array}{ll}
\Delta^{AF}(T), & \mbox{for $h_{+} > h_{-}$} \\
\Delta^{F}(T) & \mbox{for $h_{-} > h_{+}$}. \\
\end{array}
\right. \label{AF_gap}
\end{eqnarray}
This natural order parameter in a multidimensional parameter space
${\mu, U,T}$ at nonzero temperature will be called a {\sl spin
pairing pseudogap}. We define the crossover temperatures $T_s^P$
and $T_F$ as the temperatures at which the corresponding spin
pseudogaps vanish and a spin paramagnetic state, with $h_{-} =
h_{+}$, is stable. Similarly, following the peaks in {\it zero
magnetic field} spin susceptibility, a spin pseudogap
$\Delta^s(T)$, and an associated $T_s(\mu)$ can be
defined~\cite{PRB}.

\section{Results}

\subsection{$\left\langle N\right\rangle$ versus $\mu$}\label{A}

In Fig.~\ref{fig:num} for the 4-site cluster, we explicitly show
the variation of $\left\langle N\right\rangle$ {\it versus} $\mu$
below half filling for various $U$ values in order to track the
variation of charge gaps with $U$. The opening of the gap is a
local correlation effect, and clearly does not follow from long
range order, as exemplified here. For infinitesimal $U>0$, true
gaps at $\left\langle N\right\rangle=1,2$ develop and they
increase
 monotonically with $U$. In contrast, the charge gap at
$\left\langle N\right\rangle=3$ opens at
 $U\geq U_c(0)$ (a critical
value; see Ref.~\cite{PRB}) and relatively low temperatures. Thus
at low temperature, $\left\langle N\right\rangle$ (expressed as a
function of $\mu$ in Fig.~\ref{fig:num}) evolves smoothly for
$U\leq U_c(0)$, and shows finite leaps across the MH plateaus at
$\left\langle N\right\rangle=2,4$. In Fig.~\ref{fig:num}, in the
vicinity $\left\langle N\right\rangle=3$, one can notice two
phases. At $U\leq U_{c}(T)$, a negative charge gap with midgap
states is a signature of electron-electron pairing at low
temperatures. For $U\geq U_{c}(T)$, the MH-like charge gap in
(\ref{electron-hole_gap}) is positive $\mu_+>\mu_-$, which favors
electron-hole pairing similar to MH gap at half filling. As $U$
increases, $\left\langle s^z\right\rangle$ {\it versus} $\mu$
reveals islands of stability; the minimal spin state $\left\langle
s^z\right\rangle=0$ at $U\leq U_c(0)$; unsaturated ferromagnetism
$\left\langle s^z\right\rangle={1\over 2}$ at $U_c(0)<U< U_F(0)$;
saturated Nagaoka ferromagnetism, with maximum spin $\left\langle
s^z\right\rangle={3\over 2}$ at $U>U_F(0)$ (not shown in
Fig.~\ref{fig:num}).
 As $U\to \infty $,
the charge and spin gaps at $\left\langle N\right\rangle\simeq 3$
gradually saturate to its maximum $\to 2(2-\sqrt{2})t$ and
$2-\sqrt{3}$ values respectively. At $\left\langle
N\right\rangle\simeq 2$, we have full charge-spin reconciliation
when the spin gap at quarter filling approaches the charge gap,
$4(\sqrt{2}-1)t$, as $U\to \infty$. The chemical potential gets
pinned upon doping in the midgap states at $\left\langle
N\right\rangle\simeq 3$ and $U=4$. Such a density profile of
$\left\langle N\right\rangle$ {\it versus} $\mu$ near
$\left\langle N\right\rangle=3$, closely resembles the one
calculated at $U=-4$ for the {\sl attractive} 4-site Hubbard
cluster in Fig.~\ref{fig:num} and, is indicative of possible
particle or hole pairing.
\subsection{Quantum critical points}
The exact expression for the charge gap, $\Delta^{c}_{3}(U:0)$, at
$\left\langle N\right\rangle\simeq 3$ has been derived earlier in
Ref.~\cite{JMMM}. At zero temperature, the sharp transition at
critical parameter $U_c(0)$ is defined from the condition
$\Delta^{c}_{3}(U_c:0)\equiv 0$ which yields,
$U_{c}(0)=4.58399938$. The vanishing gap at $U_{c}(0)$ in the
ground state can be directly linked to the quantum critical
point~\cite{Sachdev} (QCP) for the {\sl onset of pair formation}.
Indeed the QCP, $U=U_c(0)$, separates electron-electron pairing
from MH-like electron-hole pairing regime. The QCP turns out to be
a useful point for the analysis of the phase diagram at zero and
non-zero temperatures. The QCP and the corresponding singular
doping dependencies on the chemical potential and the departure
from that point at nonzero crossover temperatures on $T-\mu$ phase
diagrams for various $U$ values are given in section~\ref{B}.
Exactly at $U=U_c(0)$ there is no charge-spin separation and, a
spin paramagnetic state coexists with a Fermi liquid similar to
non-interacting electrons, where $U=0$. At zero temperature the
analytical expression for the charge gap first was derived in
Ref.~\cite{JMMM}. Using definition introduced in
(\ref{electron-hole_gap}), the corresponding electron-electron
pairing gap, $\Delta^{P}(0)$,
\begin{eqnarray}
\Delta^{P}(0)= -{2\over \sqrt3}\sqrt{({16}+{U^2})}\cos{\gamma\over
3}-{U\over 3}+\nonumber\\ {2\over
3}\sqrt{({48}+{U^2})}\cos{\alpha\over 3}-
\sqrt{32+U^2+4\sqrt{64+3U^2}}, \label{gap_pairing}
\end{eqnarray}
exists only at at $U<U_{c}(0)$. In contrast, the electron-hole
pairing is zero at $U<U_{c}(0)$ and exists in the ground state
only for all $U>U_c(0)$. The electron-hole pairing gap,
$\Delta^{e-h}(0)$, within the range $U_{c}<U<U_{F}(0)$ is
\begin{eqnarray}
\Delta^{e-h}(0)={2\over \sqrt3}\sqrt{({16}+{U^2})}\cos{\gamma\over
3}+{U\over 3}+\nonumber\\\sqrt{32+U^2+4\sqrt{64+3U^2}}-{2\over
3}\sqrt{({48}+{U^2})}\cos{\alpha\over 3},\label{gap_el-h pairing}
\end{eqnarray}
Above the critical point $U\geq U_F(0)$ the electron-hole gap is
\begin{eqnarray}
\Delta^{e-h}(0)={2\over \sqrt3}\sqrt{({16}+{U^2})}\cos{\gamma\over
3}-\nonumber\\{2\over 3}\sqrt{({48}+{U^2})}\cos{\alpha\over 3}+
{U\over 3}+4. \label{gap_el-h}
\end{eqnarray}
where corresponding parameters in Eqs.~(\ref{gap_pairing}),
(\ref{gap_el-h pairing}) and (\ref{gap_el-h})
$\alpha=\arccos{\{({4U\over 3}-{U^3\over 27}) / ({16\over 3}
+{U^2\over 9})^{{3\over 2}}\}}$ and $\gamma=\arccos{\{(4U)/
{({16\over 3}+{U^2\over 3})^{3\over 2}}\}}$. At relatively large
$U\ge 4.584$, the energy gap $\Delta^{c}_{3}(U:0)$ becomes
positive for $\left\langle N\right\rangle=3$. With increasing
temperature, this pseudogap $\Delta^{c}_{3}(U:T)$ increases. Due
to (ground state) level crossings, the spin degeneracy in an
infinitesimal magnetic field is lifted at QCP, $U_{F}(0)=8+4\sqrt
{7}\simeq 18.583$ (\ref{gap_el-h}) and the ground state becomes a
ferromagnetic insulator with the maximum spin $\left\langle
s^z\right\rangle={3\over 2}$~\cite{PRB,Mattis}. The critical
value, $U_{c}(T)$ at which $\Delta^c_{3} (U:T)=0$, depends on the
temperature. At $U<U_{c}(0)$, $\Delta^c_{3}(T)$ in a limited range
of $U$ is negative~\cite{PRB}. Thus according to
Eq.~(\ref{charge_gap}), the states with $\left\langle
N\right\rangle=3$ become energetically less favorable when
compared with $\left\langle N\right\rangle=2$ and $\left\langle
N\right\rangle=4$ states. This is explicit evidence for the
electron phase separation instability and the existence (at finite
temperature) of particle-particle or hole-hole binding despite a
bare electron repulsion. At zero temperature, $U>U_{F}(0)$ and
$\left\langle N\right\rangle=3$, the calculated spin gap
$\Delta^s_{3}(U:T)$, for transition from $\left\langle
s^z\right\rangle={1\over 2}$ to $\left\langle
s^z\right\rangle={3\over 2}$,
\begin{eqnarray}
\Delta^{s}_{3}(U:T=0)={\sqrt{32+U^2+4\sqrt{64+3U^2}}\over
2}-2-{U\over 3},
 \label{gap_pairing1}
\end{eqnarray}
is also negative, which manifests a thermodynamic instability
$(\chi_s<0)$ and the phase separation into ferromagnetic "domains"
with $\left\langle s^z\right\rangle={3\over 2}$ and $\left\langle
s^z\right\rangle=-{3\over 2}$ in the ensemble of clusters. Thus we
can classify and see various phases, phase transitions and phase
separations in finite systems. Therefore, phase transitions in
interacting many-body system is not a characteristic of an
infinitely large thermodynamic system, but can well be defined in
canonical and grand canonical ensembles for finite systems without
the use of the thermodynamic limit.
\begin{figure} 
\begin{center}
\includegraphics*[width=20pc]{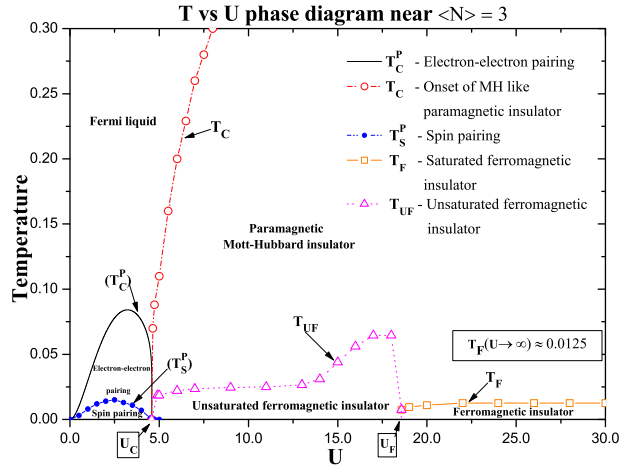}
\hfill
\end{center}
\caption {Crossover temperatures versus $U$ for 4-site phase
diagram near optimally doped $\left\langle N\right\rangle=3$
regime. Here $T_c$ denotes the crossover temperature for the onset
of the MH-like paramagnetic insulator with electron-hole pairing
gap and spin (gapless) liquid, onset of electron-electron pairing
below $T^P$, Bose condensation of electron charge and coupled spin
pairs below $T_c^P$, unsaturated ($T_{UF}$) and saturated
ferromagnetic ($T_F$) crossovers in MH-like insulators with
electron-hole pairing. Note that electron-electron pairing is
unlikely to occur above $T>0.08$ in a spin (gapless) liquid phase.
The energies and temperatures are measured in units of $t=1$, the
hopping parameter.} \label{fig:ch_gap}
\end{figure}

\subsection{Electron charge and spin pairings}

The fact that electron pairing can arise directly from the on-site
electron repulsion between electrons in small clusters was found
quite early~\cite{Bickers,scalettar,white,Cini}. However, it is
not {\it a priori} obvious that such a mechanism in an ensemble of
small clusters can survive at finite temperatures. 
  Our exact solution demonstrates that it does survive. In
Fig.~\ref{fig:ch_gap}, we show  condensation with bound, double
electron charge and decoupled spin at $U\leq U_c(0)$ and $T_s^P(U)
\leq T \leq T_c^P(U)$. Below $T_s^P(U)$,
 the spin degrees are also condensed
with 'bosonization' of spin and charge degrees and possible
'superconductivity'. As an important observation we found is that
the variation of pairing gaps, $\Delta^{P}(U)$ and
$\Delta^{e-h}(U)$ {\it versus} $U$, in the ground state closely
follows the variation of $T_c^P(U)$, and $T_c(U)$ respectively.
For $U\leq U_c(0)$ there is correlation between the spin pairing
gap $\Delta_s^P(U)$ and corresponding crossover temperature
$T_s^P(U)$ {\it versus} $U$. At larger $U\geq U_c(0)$, we found
similar correlations in $U$ space between ferromagnetic spin
pairing gaps and corresponding ferromagnetic crossover
temperatures.

Our calculations for the $U\leq U_c$ range may also be used to
reproduce the behavior of T$_c (p)$ versus pressure $p$ in the
HTSCs, if we assume that the parameter $U$ decreases with
increasing pressure. In Fig.~\ref{fig:ch_gap}, the crossover
temperatures for electron-electron, $T_c^P$, electron-hole, $T_c$,
and, coupled spin, $T_s^P$, pairings are plotted as a function of
$U$. The shown condensation of both doubled electron charge, and
zero spin $s^z=0$ (singlet) degrees below $T_s^P$ is indicative of
a possible mechanism of superconductivity in the HTSCs near
optimal doping. In the HTSCs, superconducting transition
temperature
 T$_c$ generally increases with pressure first,
reaches the maximum value at some critical pressure $p_c$ and then
decreases with pressure~\cite{Xiao} in agreement with our result
for the variation of $T_s^P(U)$ {\it versus} $U$. These pressure
effects in the cuprate family reflect the changes in electron
pairing due to the moderate Coulomb interaction $U$.
Notice, that at enough large $U$, $T_c(p)$ decreases with $U$
under pressure, as it is shown in Fig.~\ref{fig:ch_gap}. This
might explain why the pressure $T_{c}(p)$ decreases across some of
organic and families alkali doped fullerene
superconductors~\cite{Zheng}.

Our results for $N\approx 3$ in Fig.~\ref{fig:ch_gap} suggest that
the enhancement of T$_c$ in the parental MH cuprates, with
relatively large $U$ in the optimally doped HTSCs, can be due to
an increase of pairing by decreasing $U$ under pressure rather
than an increase of the pressure-induced hole concentration. Thus
it appears that the 4-site cluster near $N\approx 3$ indeed
captures the essential physics of the pressure effect on electron
pairing and BC in HTSCs near optimal doping. Similarly, our
calculations in Fig. \ref{fig:ch_gap} for $U\geq U_c$ range can
also be useful in predicting the variation of T$_{UF} (p)$ and
T$_F (p)$ versus pressure $p$ in ferromagnetic insulators and
Co-doped anatase TiO$_2$~\cite{Yuji}.

\subsection{Charge-spin separation}

Until recently, electrons were thought to carry their charge and
spin degrees simultaneously. It is certainly true for
non-interacting electrons, where $U=0$. At zero temperature, a
paramagnetic Fermi liquid with zero charge and spin gaps in
Fig.~\ref{fig:ch_gap} exists only at $U=0$ and $U=U_c(0)$. At
finite temperatures the corresponding charge pseudogaps vanish at
particle-partricle/hole pairing crossover temperatures and there
is also no
 sign of charge and spin separation above $T_c$ and
$T_c^P$. However, as temperature decreases below these crossover
temperatures only the charge pseudogaps are formed, while spin
excitations
 remain gapless and these paramagnetic spin
liquid states coexisting with electron-electron or MH-like
electron-hole pairings will be labeled as a charge-spin
separation. Gapless spin response due to small variations of
magnetic field or electron concentration (chemical potential) is
independent of the charge degrees of freedom. By decreasing
temperature, a partial charge-spin reconciliation takes place in
Fig.~\ref{fig:ch_gap} below ferromagnetic crossover temperatures
due to the formed spin pseudogap with coupled charge and spin
degrees. Accordingly, electron-hole pairs and ferromagnetic
coupled spins are bounded in charge and spin sectors below
ferromagnetic crossover temperature. The effect of charge-spin
separation becomes stronger with increasing $U>U_c(0)$, since the
distinction between the corresponding charge and spin crossover
temperatures increases with $U$ (see Fig.~\ref{fig:ch_gap}).
\begin{figure} 
\begin{center}
\includegraphics*[width=20pc]{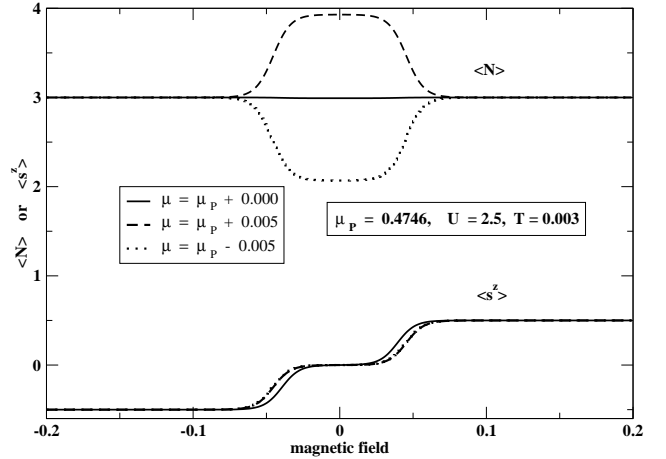}
\hfill
\end{center}
\caption {The average number of electrons and magnetization
dependencies versus an applied magnetic field at $U=2.5$,
$T=0.003$ and $\mu_P=0.4746$. Note how a small change in the
chemical potential (from $\mu_P\pm 0.005$) forces the (paired)
system into a magnetic spin liquid state with an unpaired spin and
$N \approx 3$. } \label{fig:degree_freedom_1}
\end{figure}

In Fig.~\ref{fig:degree_freedom_1}, we examine the behavior of
charge and spin degrees of freedom below the spin condensation
temperature $T_s^P$. Fig.~\ref{fig:degree_freedom_1} shows how the
spin degrees, at an infinitesimal magnetic field, follow the
charge. Notice how strong inter-configuration charge and spin
fluctuations can be monitored by small variation of a chemical
potential or magnetic field. The coupling between the spin and
charge degrees of freedom is manifested due to the composite
nature of the electron, having charge (holon) and spin (spinon).
Thus, the spin singlet pairing in the spin sector and the
electron-electron or hole-hole pairings in the charge sector are
not independent but demonstrate coherency and a strong coupling
between them, which is necessary for possible 'superconductivity'.
Below the critical temperature of crossover $T_c^P$, the charge
degrees are coupled and simultaneous condensation of spin degrees
below $T_s^P$ results in reconciliation of charge and spin and
possible full 'bosonization' of electrons.
Thus the electron fragments into the charge and spin excitation
states and superconductivity arises due to the charge-spin
separation and subsequent condensation of these charge and spin
degrees that are both bosonic in nature~\cite{RVB}.

\subsection{Phase separation}

The mechanism of phase separation (i.e segregation) in small
clusters near $\left\langle N\right\rangle=3$ is also quite
straightforward and simply depends on the pairing conditions in
charge and spin channels respectively~\cite{PRB}. At $U>U_c(0)$ in
Fig.~\ref{fig:ch_gap}, we notice a stable ferromagnetic state with
unsaturated magnetization and a ferromagnetic state with saturated
Nagaoka magnetization at larger $U>U_F(0)$ values. As temperature
increases the system undergoes a smooth crossover from a
ferromagnetic MH-like insulator with $\Delta^{e-h}\neq 0$ and
$\Delta_F\neq 0$ into a paramagnetic MH insulator with
$\left\langle s^z\right\rangle\approx {0}$. The ferromagnetic
critical temperature $T_F (U)$ increases monotonically with $U$
and as $U\to \infty$ the limiting $T_F (U \to \infty)=0.0125$
value approaches to the maximum spin pairing temperature $T_s^P$.
A ferromagnetic phase with broken symmetry was obtained here in
the presence of an infinitesimal magnetic field and increasing
temperature leads to a F-PM transition and restoration of the
symmetry in paramagnetic phase. The level crossing at $U_c(0)$
$(U_F(0))$ in the presence of an appropriate infinitesimal
magnetic field brings about phase separation of the
thermodynamically unstable ensemble of clusters with $h_{-}>h_{+}$
into spin up and spin down ferromagnetic "domains" (see
section~\ref{A}).

\begin{figure} 
\begin{center}
\includegraphics*[width=20pc]{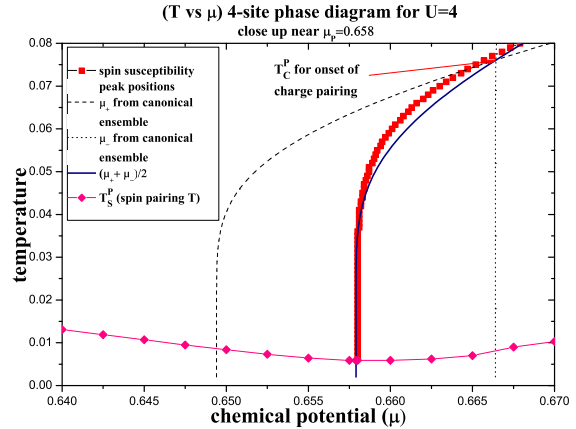}
\end{center}
\caption {The phase $T$-$\mu$ diagram in canonical ensemble of
four-site clusters close to $\left\langle N\right\rangle\approx 3$
and $\mu_P=0.658$ at $U=4$. The phase below $T_{c}^{P}$ suggests
the existence of electron-electron pairing at finite temperature
with unpaired spin states. The zero (magnetic) spin susceptibility
peak in spin (melted) liquid state terminates at finite
$T_{s}^{P}$ and as temperature is further lowered the spin pairs
begin to form. This picture with $\mu_{-}(T)>\mu_{+}(T)$ supports
the idea that there is inhomogeneous, electronic phase separation
here. As $U$ increases above $U_c(0)=4.584$, these inhomogeneities
disappear and a stable unsaturated ferromagnetic MH-like
insulating region emerges around optimal doping shown in
Fig.~\ref{fig:phase_u6}.} \label{fig:phase_u4}
\end{figure}

It appears that the canonical approach yields also an adequate
estimation of possible pair binding instability near $\left\langle
N\right\rangle=3$ in an ensemble of small clusters at relatively
low temperature and moderate $U\leq U_c(0)$. New important
features appear if the number of electrons $\left\langle
N\right\rangle=3$ is kept fixed for the whole system of decoupled
clusters, placed in the (particle) bath, by allowing the particle
number on each separate cluster to fluctuate. One is tempted to
think that due to symmetry, there is a single hole on each cluster
within the $\left\langle N\right\rangle=3$ set in the  ensemble.
However, due to thermal and quantum fluctuations in the density of
holes between the clusters $U<U_c(0)$), it is energetically more
favorable to form pairs of holes. In this case, snapshots of the
system at relatively low temperatures and at a critical value
($\mu_P$ in Fig.~\ref{fig:phase_u4}) of the chemical potential
would reveal equal probabilities of finding (only) clusters that
are either hole rich ($\left\langle N\right\rangle=2$) or hole
poor ($\left\langle N\right\rangle=4$). Thus ensemble of 4-site
clusters at $\left\langle N\right\rangle=3$ is thermodynamically
unstable, $\mu_{+}\leq \mu_{-}$, and can lead to macroscopic phase
separation onto hole-rich and hole-poor regions recently detected
in super-oxygenated La$_{2-x}$Sr$_x$CuO$_{4+y}$, with various Sr
contents~\cite{hashini,PRB}.

The level crossing under slight doping (infinitesimal variation of
chemical potential) brings about phase separation of ensemble of
clusters and this can be linked to the formation of inhomogeneity,
consisting of hole-poor (superconducting) and hole-rich
(antiferromagnetic) "domains" at $U<U_c(0)$. Thus we can conclude
that ensemble of 4-site clusters with $\left\langle
N\right\rangle=3$  at relatively low temperatures is unstable for
all $U>0$ values with regard to phase separation due to
spontaneous symmetry braking into ferromagnetic or superconducting
"domains" either in spin or charge sectors respectively.

The $T$-$\mu$ phase diagram for the 4-site cluster near $\mu_P$,
is shown in Fig.~\ref{fig:phase_u4}. This exact phase diagram at
$U=4$ in the vicinity of the optimally doped ($N\approx 3$) regime
has been constructed based on the condition ${\Delta^{c}}(T)<0$
with $\mu_{-}(T)>\mu_{+}(T)$, reported earlier in Ref.~\cite{PRB}.
The electron pairing temperature, $T_{c}^{P}$, identifies the
onset of charge pairing. As temperature is further lowered, spin
pairs begin to form at $T_{s}^{P}$. At this temperature (with zero
magnetic field), spin susceptibilities become very weak indicating
the disappearance of the $\left\langle N\right\rangle\approx 3$
states. Below this spin pairing temperature, only paired states in
"metallic" liquid ($\Delta^{e-h}=0$) in the midgap region are
observed to exist having a certain rigidity, so that a nonzero
magnetic field or a finite temperature is required to break the
pairs. From a detailed analysis, it becomes evident that the
system is on the verge of an instability; the paired phase
competing with a phase that suppresses pairing which has a high,
zero-field magnetic susceptibility. As the temperature is lowered,
the number of $\left\langle N\right\rangle\approx 3$ (unpaired)
clusters begins to decrease
 while a mixture of (paired)
 $\left\langle N\right\rangle\approx 2$ and
 $\left\langle N\right\rangle\approx 4$
 clusters appears. In Fig.~\ref{fig:phase_u4}
 the spin pairing phase below $T_{s}^{P}$ competes with a phase (having a high
magnetic susceptibility) that suppresses pairing at `moderate'
temperatures.
  Surprisingly, the
  critical doping  $\mu_P$
 (which corresponds to a filling
 factor of $1/8$ hole-doping away from half filling),
 where the above pairing
 fluctuations take place when
 $U<U_c(0)$,
  is close to the doping level near which numerous intriguing
 properties have been observed
 in the hole-doped HTSCs. For example, the spin
pseudogap can be driven to zero also by applying a suitable
magnetic field. This factor leads to the stability of electron
"dormant" magnetic configuration in a narrow, critical doping
region close to $\left\langle N\right\rangle\approx 3$, competing
with $\left\langle N\right\rangle=2$ and $\left\langle
N\right\rangle=4$ states as have seen in a recent
experiment~\cite{hashini}. Thus our results, at electron
concentration $\left\langle N\right\rangle\approx 3$, clearly
demonstrate phase separation and breakdown of Fermi liquid
behavior due to electron-electron/hole pairings and Nagaoka
ferromagnetism in the absence of long-range order.

\subsection{Phase diagrams}\label{B}

\begin{figure} 
\begin{center}
\includegraphics*[width=20pc]{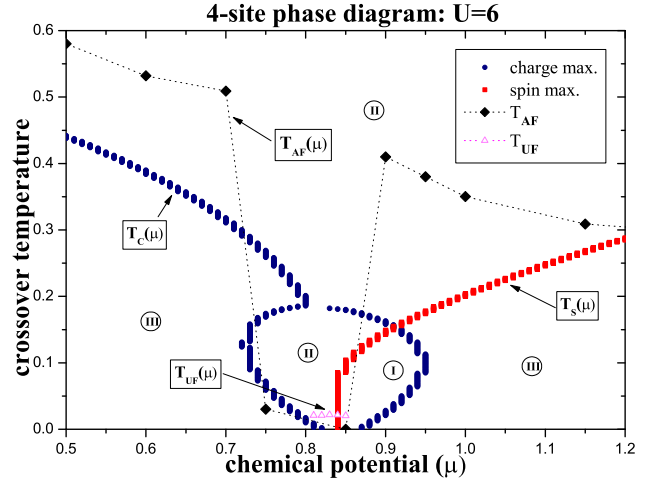}
\end{center}
\caption {Temperature $T$ vs chemical potential $\mu$ phase
diagram
 for the grand canonical ensemble of four-site clusters at $U=6$ and $h=0$.
Regions I, II and III are quite similar to the ones found in
enlarged Fig.~\ref{fig:phase_u4} for $U=4$~\cite{PRB}, again
showing strong charge-spin separation.  However, a charge gap
opens as a new bifurcation (I and II phases) which consists of
charge and spin pseudogaps, (replacing the line phase P in the
previous figure: see text). In the the vicinity of $\left\langle
N\right\rangle\approx 3$ below $T_{UF}(\mu)$ unsaturated
ferromagnetism coexists with the MH-like insulator.}
\label{fig:phase_u6}
\end{figure}

In Figs. \ref{fig:phase_u4} and \ref{fig:phase_u6}, the phase
diagrams for the ensemble of 4-site clusters are shown, where we
define the charge peak (i.e maxima) $T_c(\mu)$ to be the
temperature with maximum $\chi_c(\mu)$ at a given $\mu$. Possible
peaks in the zero magnetic field spin susceptibility
$\chi_s(\mu)$, when monitored as a function of $\mu$, can also be
used to define an associated temperature, $T_s(\mu)$.
 In addition,
 for a given $\mu$, $T_{AF}(\mu)$
defines the temperature at which AF (spin) gap disappears, i.e.
${\Delta^{s}}(T_{AF})=0$. These have been constructed almost
exclusively using the temperatures, $T_c(\mu)$, $T_s(\mu)$ and
$T_{AF}(\mu)$, defined previously. We have identified the
following phases in these diagrams: (I) and (II) are charge
pseudogap phases separated by a phase boundary where the spin
susceptibility reaches a maximum, with $\Delta^{e-h}(T)>0$,
$\Delta^{AF}(T)=0$; at finite temperature, phase I has a higher
$\left\langle N\right\rangle$ and coupled spin compared to phase
II spin liquid phase; Phase (III) is a MH-like antiferromagnetic
insulator with bound charge and spin, when $\Delta^{e-h}(T)>0$,
$\Delta^{AF}(T)>0$; phase separation (PS) in $T-\mu$ plane for
$U=4$ with a vanished charge gap at $\left\langle
N\right\rangle=3$, now corresponding to the opening of a pairing
gap ($\Delta^P(T)>0$) in the electron-electron channel with
$\Delta^c _{3} (U: T)<0$. We have also verified the well known
fact that the low temperature behavior in the vicinity of half
filling, with charge and spin pseudogap phases coexisting,
represents an AF insulator ~\cite{JMMM}. However, {\sl away from
half filling}, we find very intriguing behavior in thermodynamical
charge and spin degrees of freedom. In both phase diagrams, we
find similar MH-like (I), (II) and AF-like (III) charge-spin
separated phases in the hole-doped regime. In
Fig.~\ref{fig:phase_u6}, spin-charge separation in phases (I) and
(II) originates for relatively large $U$ in the underdoped regime.
 In contrast, Fig.~\ref{fig:phase_u4}
shows the existence (at $U=3$) of a line phase (with pairing)
similar to $U<0$ case with electron pairing ($\Delta^P(T)>0$),
when the chemical potential is pinned up on doping within the
highly degenerate midgap states near (underdoped) $1/8$ filling.

Among other interesting results, rich in variety for $U>0$, sharp
transitions and quantum critical points (QCPs) are found between
phases with true charge and spin gaps in the ground state; for
infinitesimal $T>0$, these gaps are transformed into `pseudogaps'
with some nonzero weight between peaks (or maxima) in
susceptibilities monitored as a function of doping (i.e. $\mu$) as
well as $h$. We have also verified the well known fact that the
low temperature behavior in the vicinity of half filling, with
charge and spin pseudogap phases coexisting, represents an AF
insulator in the Hubbard clusters~\cite{JMMM}. However, {\sl away
from half filling}, we find very intriguing behavior in
thermodynamical charge and spin degrees of freedom. In both phase
diagrams, we find similar MH-like (I), (II) and AF-like (III)
charge-spin separated phases in the  hole-doped regime. In
Fig.~\ref{fig:phase_u6},
 spin-charge separation in
phases (I) and (II) originates for relatively large $U$ in the
underdoped regime. We have also seen that a reasonably strong
magnetic field has a dramatic effect (mainly) on the QCP at
$\mu_s$, at which the spin pseudogap disappears. The regions (I)
and (II), with the prevailing charge gap $\Delta^c(T)>0$, are
separated by a boundary where the spin gap vanishes. At critical
doping $\mu_s$ with $T_s \to 0$, the zero spin susceptibility
$\chi_s$ exhibits a sharp maximum. The critical temperature $T_s
(\mu)$, which falls abruptly to zero at critical doping $ \mu_s$,
implies~\cite{Tallon} that the pseudogap can exist independently
of possible superconducting pairing in Fig.~\ref{fig:phase_u6}. In
contrast, Fig.~\ref{fig:phase_u4} shows the existence at $U=$ 3 of
a line phase (with pairing) similar to $U<0$ case with a spin
pseudogap ($\Delta^s(T)>0$) and electron pairing pseudogap
($\Delta^P(T)>0$), when the chemical potential is pinned up on
doping within the highly degenerate midgap states near
(underdoped) $1/8$ filling.

We have also seen that a reasonably strong magnetic field has a
dramatic effect (mainly) on the QCP at $\mu_s$, at which the spin
pseudogap disappears. It is evident from our exact results that
the presence of QCP at zero temperature and critical crossover
temperatures, give strong support for cooperative character of
existing phase transitions and crossovers in finite size clusters
as in large thermodynamic systems~\cite{Langer,Cyrot}.

As an important remark, in the noninteracting case, $U=0$, the
charge and spin peaks follow one another (in sharp contrast to the
$U=$ 4 and 6 cases, in regions I and II where charge (as well as
spin) maxima and minima are well separated) indicating that there
is no charge-spin separation, even in the presence of a magnetic
field. In the entire range of $\mu$, the charge and spin
fluctuations directly follow one another, i.e. without charge-spin
separation. For interacting electrons, we notice similar respond
of charge and spin degrees at $\left\langle N\right\rangle=3$ only
at single point, $U=U_c(0)$. In the atomic limit, $t=0$, a full
charge-spin separation, exactly at half filling, takes place in
the ground state and the corresponding MH crossover temperature of
the metal-insulator transition occurs at $T_{MH}={U/(2\ln{2})}$.
\begin{figure} 
\begin{center}
\includegraphics*[width=20pc]{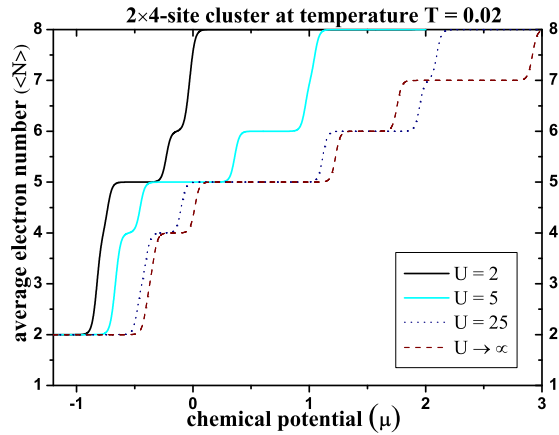}
\hfill
\end{center}
\caption {Variation of average electron concentration versus $\mu$
in ensemble of ${2\times 4}$ clusters with the couplings $c=1$
between the squares for various $U$ values, $h=0$ and $T=0.02$.
Here we can easily identify regions quite similar to the ones
found in Fig.~\ref{fig:num}, showing strong charge-spin separation
with various crossover temperatures. This phase suggests the
existence of electron-electron pairing at low temperatures, $T\leq
0.028$.}
 \label{fig:gap_1 and 2}
\end{figure}
\subsection{Coupled 4-site clusters}
We have also carried out exact numerical diagonalization and
calculations of the charge gap and pairing in 8-site, $2\times 4$
planar clusters with periodic boundary conditions in both
directions, to illustrate similar effects on the properties
described above for the 4-site clusters. The pairing fluctuations
that are seen for the 4-site cluster are found to exist for
${2\times 4}$ ladders near half filling ($\left\langle
N\right\rangle\approx$ 7) as well. Most of the trends observed for
the 4-site clusters, such as the MH-like charge (pseudo) gap,
AF-like (pseudo) gap, (spin) pseudogap, electron and spin pairing
(pseudo) gaps and with corresponding crossover temperatures are
also observed here. Moreover, we find corresponding critical
$U_c(0)$ and $U_F(0)$ values for the pairing instabilities and
vanishing of corresponding pseudogaps at finite temperatures
$T_{UF}$ and $T_F$ similar to the 4-site cluster. The fluctuations
that occur here at optimal doping are among the states with
$\left\langle N\right\rangle\approx$ 6, 7 and 8 electrons. At
$\left\langle N\right\rangle\approx$ 7, we clearly observe the
dormant magnetic state as  noticed in the 4-site cluster with a
slight variation of the chemical potential or magnetic field. Thus
our exact cluster simulations of the Hubbard model displays
incipient pairing, ferromagnetism, phase separation and other
phenomena are remarkably similar to those found in small
nanoparticles, transition metal oxides and high T$_c$
cuprates~\cite{Nagaev}.

\section{Conclusion}

In summary, we have illustrated how to obtain phase diagrams and
identify the presence of temperature driven particle-particle/hole
and spin pairing
crossovers below $U< U_c(0)$, quantum critical points ($\mu_s$,
$\mu_c$) and charge-spin separation regions for any $U>U_c(0)$ in
the ensemble of 4-site Hubbard {\it clusters} as doping (or
chemical potential) is varied. Specifically, our exact solution
pointed out an important difference between the $U=4$ and $U=6$
phase diagrams near half filling (i.e. one hole off half filling),
which can be tied to electron-electron pairing and possible
superconductivity and ferromagnetism in doped HTSCs and transition
metal oxides or disulfides respectively. Our results show the
pairing crossover near $\left\langle N\right\rangle\approx 3$ and
strongly suggest that particle-particle pairing and spin coupling
can exist at $U<U_c(0)$, while particle-hole binding and
ferromagnetism is presumed to occur for $U> U_c(0)$. Exactly at
$U=U_c$, paramagnetic Fermi liquid is stable in the ground state
and all finite temperatures. At $U> U_F(0)$, there is another
subsequent transition into a saturated ferromagnetic insulator
with maximum spin $\left\langle s^z\right\rangle\approx {3\over
2}$. It is also apparent that short-range correlations alone are
sufficient for pseudogaps to form in small 4-site and larger
${2\times 4}$ clusters, which can be linked to the generic
features of phase diagrams in temperature and doping effects seen
in the HTSCs. The exact cluster solution shows how charge and spin
gaps are formed at the microscopic level and their behavior as a
function of doping (i.e. chemical potential), magnetic field and
temperature. The spin and charge crossover temperatures can also
be associated with the formation of pairing gaps below $T_c^P(U)$.
As temperature decreases further, a simultaneous BC of spin
degrees of freedom takes place below $T_s^P(U)$. The increase of
$T_s^P(U)$ with decrease of $U$ below $U_{c}$ in
Fig.~\ref{fig:ch_gap} reproduces the variation of T$_c$ {\it
versus} $p$ in the optimally and nearly optimally doped HTSC
materials~\cite{Xiao}, which suggests a significant increase of
pairing temperature T$_c$ under pressure or $U$ due to the
enhancement of the zero temperature charge and spin pairing gaps.
In addition, our calculations provide important benchmarks for
comparison with Monte Carlo, RSRG, PCT and other approximations.

Finally, we have developed novel theoretical concepts and
techniques, which are especially suitable and efficient for
understanding of nascent superconductivity and magnetism using the
canonical and grand canonical ensemble for small clusters. Our
results show the cooperative {\it nature} of phase transition
phenomena in finite-size clusters similar to large thermodynamic
systems. The developed approach allows an exact and unbiased study
of the Fermi liquid instabilities in small clusters without the
assumption of a broken symmetry. The small {\it nanoclusters}
exhibit particle-particle, spin-spin pairings in a limited range
of $U$, $\mu$ and $T$ and share very important intrinsic
characteristics with the HTSCs and magnetic oxides. These ideas
could be useful in different areas outside the cluster field, to
systems ranging from molecules to continuous media, and applied
for understanding of phase separation and incipient spontaneous
superconductivity and ferromagnetism in small nanometer-scale
clusters and Nb, Co
nanoparticles~\cite{deHeer0,deHeer1,Tinkham,Cox}. This research
was supported in part by the U.S. Department of Energy under
Contract No. DE-AC02-98CH10886.

\end{document}